\documentclass[twoside]{dis04}
\def\runauthor{L.Favart}
\def\shorttitle{Deeply Virtual Compton Scattering at HERA}

\newcommand{\pbinv}{\mbox{${\rm pb^{-1}}$}}
\newcommand{\qsq}{\mbox{$Q^2$}}

\def\Journal#1#2#3#4{{#1} {\bf #2} (#3) #4}

\def\PLB{{\em Phys. Lett.} {\bf B}}

\def\PRD{{\em Phys. Rev.} {\bf D}}

\def\EPJ{{\em Eur. Phys. J.} {\bf C}}

\begin{document}

\title{Deeply Virtual Compton Scattering at HERA}

\author{Laurent Favart}

\address{{\rm On behalf of the H1 Collaboration} \\
I.I.H.E., Universit\'e Libre de Bruxelles, Belgium\\
E-mail: lfavart@ulb.ac.be}

\maketitle

\abstracts{
The Deeply Virtual Compton Scattering (DVCS)
$\gamma^* p \rightarrow \gamma p$ cross section has been measured
with the H1 detector at HERA with an increased precision and 
in an extended kinematic domain:
at photon virtualities $4<Q^2<80$ GeV$^2$, and photon-proton
c.m.s. energy $30<W<140$ GeV.
The measurement is compared to NLO QCD calculations
and to Colour Dipole model predictions.
}

\section{Introduction}
\begin{figure}[htb]
 \begin{center}
  \epsfig{figure=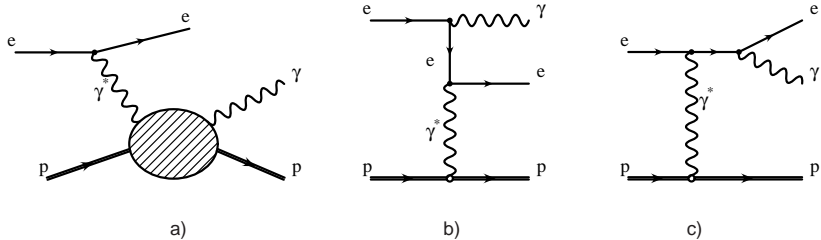,width=0.9\textwidth}
  \caption{\sl The DVCS (a) and the Bethe-Heitler (b and c) processes.}
  \label{fig:dvcs}
  \label{fig:bh}
 \end{center}
\end{figure}

Deeply Virtual Compton Scattering (DVCS), sketched in
Fig.~\ref{fig:dvcs}a,
consists of the hard diffractive
scattering of a virtual photon off a proton.
It contributes to the reaction $e^+p \rightarrow e^+ \gamma p$
as the purely electromagnetic Bethe-Heitler (BH) process
(Figs.~\ref{fig:bh}b and c) and the interference between the two
processes.
%Previous DVCS measurements at HERA can be found in
%references~\cite{h1-dvcs, zeus-dvcs, hermes-dvcs}.
\\
 
The interest of the DVCS process resides in the
particular insight it gives to the applicability of
perturbative Quantum Chromo Dynamics (QCD) in the field of
diffractive interactions.
% parallel with light VM
%%%%%%%%%%%%%%%%%%%%%%%%
%
In the presence of a hard scale, the DVCS scattering amplitude
factorises into a hard scattering part calculable in perturbative QCD
and parton distributions which contain the non-perturbative
effects due to the proton structure.
The DVCS process is similar to diffractive vector meson
electroproduction but with a real photon replacing the final state
vector meson. It avoids the
theoretical complications and uncertainties associated with
the unknown vector meson wave function.
However, even at photon virtualities $Q^2$ values above a
few GeV$^2$, non perturbative effects can still take place.
The wide kinematic range in the photon virtuality, \qsq,
accessible at HERA, provides a powerful probe for the interplay
between the perturbative and non-perturbative regimes in QCD.
Furthermore the DVCS process gives access to the Generalised Parton
Distributions (GPD) which are generalisation and unification of the familiar
parton distributions and form factors, and also includes parton momentum 
correlations.
\\

%%%%%%%%%%%%%%%%%%% Fig 1 %%%%%%%%%%%%%%%%%%%%%%%%%%%%%%%%%
%%%%%%%%%%%%%%%%%%%%%%%%%%%%%%%%%%%%%%%%%%%%%%%%%%%%%%%%%%%

At high energy, first cross section measurements of the DVCS process 
were published by H1~\cite{h1-dvcs} and ZEUS~\cite{zeus-dvcs}. 
Here, the new H1 measurement is reported, in an extended kinematic
range: $4<Q^2<80$ GeV$^2$, $30<W<140$ GeV and $|t|<1$ GeV$^2$, 
where $t$ is the squared 4 momentum transfer between the incoming and
the scattered protons. The analysis relies on the
integrated luminosity of 26 $\pbinv$ of data taken during the year 2000 
(i.e. 3.5 times larger than the previously published by
H1~\cite{h1-dvcs}). More details on the present analysis can be found in
\cite{h1eps03}.

\section{Analysis strategy}

At these small values of $t$ the reaction
$e p \rightarrow e \gamma p\,$
is dominated by the purely electromagnetic
BH process whose cross section,
depending only on QED calculations and proton elastic form factors, is
precisely known and therefore can be subtracted.
To enhance the ratio of selected DVCS
events to BH events the outgoing photon is selected in the
forward, or outgoing proton, region with transverse momentum larger than 
$2\,{\rm GeV}$. Large values of
the incoming photon virtuality $Q^{2}$ are selected by detecting the
scattered
electron in the SPACAL calorimeter with energy larger than $15\,{\rm
GeV}$. The outgoing proton escapes down the
beam-pipe in the forward direction. 
In order to reject inelastic and proton dissociation events,
no further cluster in the calorimeters with energy above
noise level is allowed and an absence of activity 
in forward detectors is required.
\\

Nevertheless an important contamination subsists from the
DVCS process with proton dissociation:
$ e^+ + p \rightarrow e^+ + \gamma + Y ,$
when the decay products of the baryonic system $Y$ are not detected in
the forward detectors.
This typically occurs for $Y$ systems with masses below 1.6 GeV.
The sum of DVCS and BH contributions, in which the proton does not
survive intact, has been estimated to be $11\pm 6\,\%$ of the final
sample. This uncertainty constitutes the main systematic error on the
measurement, with the 7\% uncertainty due to acceptance and bin center
corrections.

\section{Results}

\begin{figure}[htbp]
 \begin{center}
  \epsfig{file=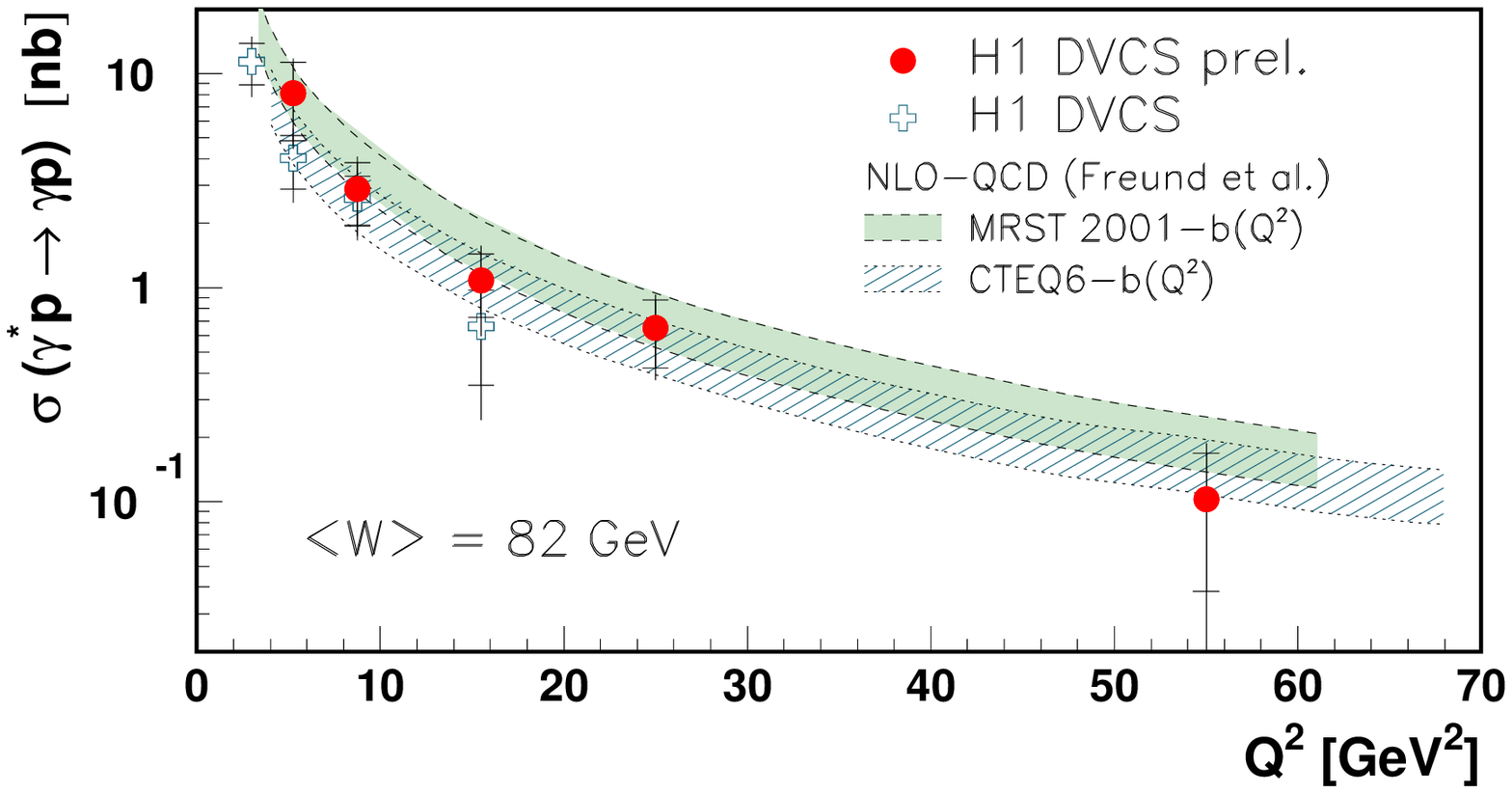,width=0.70\textwidth}
  \epsfig{file=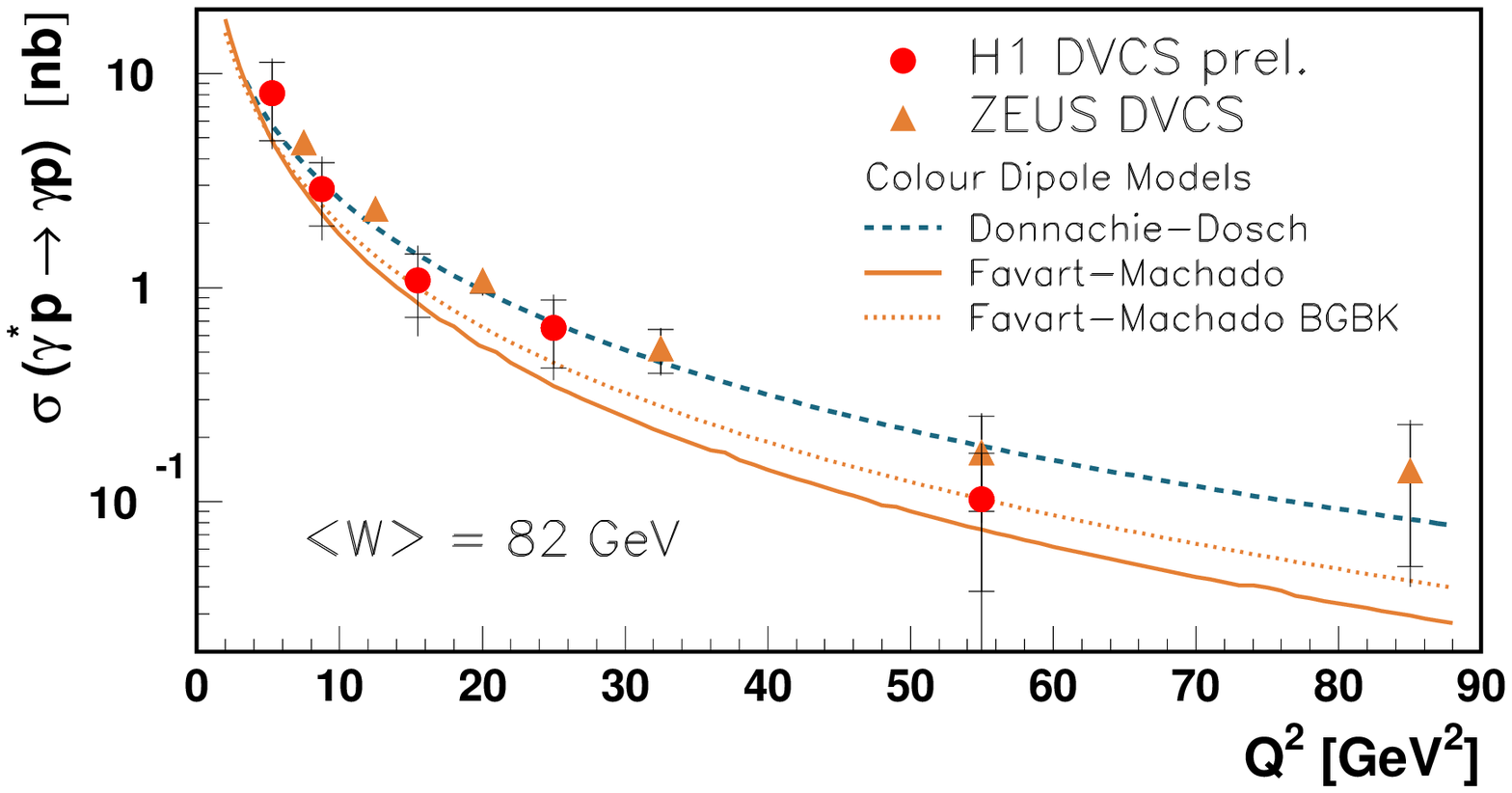,width=0.70\textwidth}
 \end{center}
 \caption[*]{$\gamma^* p \rightarrow \gamma p$
 cross section as a function of $Q^2$ for $<W>=82$~GeV.
 The inner error bars are statistical and
 the full error bars include the systematic errors added in quadrature.
 The measurement is compared (upper plot)
 to the NLO QCD prediction~\cite{Freund:2001hm,Freund:2001hd} 
 using 
 GPD parametrisations based on MRST2001 and CTEQ6~\cite{Freund:2002qf}
 The bands correspond to $b_0$ values between 5 and 9~GeV$^{-2}$.
 On the lower plot, the measurement is compared to 
 two different Colour Dipole models predictions, by Donnachie and
 Dosch~\cite{Donnachie:2000px} and by Favart, Machado~\cite{Favart:2003cu}
 at the fixed value $b=7$ GeV$^{-2}$. The BGBK notation indicates
 the additional DGLAP evolution of the dipole cross section 
 added to the basic prediction.}
\label{fig:gpsigq2}
\end{figure}

\begin{figure}[tbh]
 \begin{center}
  \epsfig{file=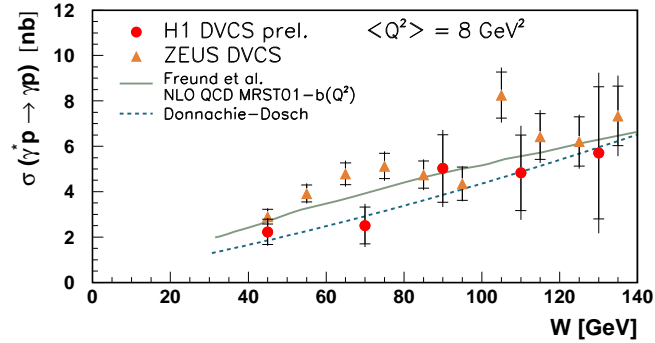,width=0.70\textwidth}
 \end{center}
 \caption[*]{$\gamma^* p \rightarrow \gamma p$
 cross section as a function of $W$ for $<Q^2>=8$~GeV$^2$.
 The measurement is compared
 to the NLO QCD prediction~\cite{Freund:2001hm,Freund:2001hd}
 using GPD parametrisation based on MRST2001~\cite{Freund:2002qf}
 and to the Colour Dipole models prediction of Donnachie and
 Dosch~\cite{Donnachie:2000px}.
}
\label{fig:gpsigw}
\end{figure}

The $\gamma^* p$ cross section for the DVCS process is shown in
Fig.~\ref{fig:gpsigq2} as a function
of $Q^2$ for $W=82$~GeV, 
and in Fig.~\ref{fig:gpsigw} as a function of $W$ for $Q^2=8$~GeV$^2$.
In the upper plot of Fig.~\ref{fig:gpsigq2} the measurement is compared
to the NLO QCD prediction~\cite{Freund:2001hm,Freund:2001hd} using two different
GPD parametrisations~\cite{Freund:2002qf}.
The  $t$ dependence is parametrised as $e^{-b|t|}$, with $b=b_0(1-0.15
\log(Q^2/2))$~GeV$^{-2}$. 
The classic PDF $q(x,\mu^2)$ of MRST2001 and CTEQ6 are used in the DGLAP region
($x>\xi$) such that $\cal H$, which is the only important GPD at small $x$ 
is given at the scale $\mu$ by:
${\cal H}^q(x,\xi,t;\mu^2)=q(x;\mu^2) \, e^{-b|t|}$ for quark singlet and
$ {\cal H}^g(x,\xi,t;\mu^2)=x\ g(x;\mu^2) \, e^{-b|t|}$ for gluons,
i.e. independent of the skewing parameter $\xi$.
Keeping this parametrisation in the ERBL region ($|x|<\xi$) would lead 
to a prediction overshooting the data by a factor 4-5.
Therefore, a parametrisation is proposed by the
authors to suppress the region of very small $x$ (for details
see~\cite{Freund:2002qf}). This emphasises the interesting sensitivity
to the ERBL region.
The NLO QCD predictions are in good
agreement with the data, for both GPD parametrisations. Since the main
difference between the two parametrisations resulting in the
normalisation, it emphasizes the need for a direct $t$ dependence
measurement. 
%It should be mentionned that these GPD model correspond to
%no external skewedness except the fact that one has to compare $q(x)$
%with $q(x_{Bj})$ and $x<<x_{Bj}$. 
\\

In the lower plots of Fig.~\ref{fig:gpsigq2} the measurement is compared
to two different Colour Dipole models predictions, by Donnachie and
Dosch~\cite{Donnachie:2000px} and by Favart and Machado~\cite{Favart:2003cu}.
They are based on a factorisation into the incoming photon
wave function, a $q\bar q$-p cross section and the outgoing photon wave
function. The models differ in the way the quark
dipole cross section is parametrised. Donnachie and
Dosch~\cite{Donnachie:2000px}
basically connect a soft Pomeron with large dipole size and a hard
Pomeron with
small dipole size. Favart and Machado~\cite{Favart:2003cu} apply the
saturation model
of Golec-Biernat et al.~\cite{Golec-Biernat:1999qd} to the DVCS process,
with a possible DGLAP evolution~\cite{Favart:2004uv,Bartels:2002es} 
quoted BGBK on the plot.
In both cases an exponential $t$-dependence, $e^{-b|t|}$, is
assumed.
All presented Colour Dipole model predictions describe well the data in
shape and in normalisation for the same value of $b=7$ GeV$^{-2}$.
\\

The new measurement is also compared to the previous
measurement by H1~\cite{h1-dvcs} 
and to the ZEUS measurement~\cite{zeus-dvcs}.
The two H1 measurements are in good agreement. The new H1 measurement
is in fair agreement with ZEUS results except for $W\sim70$ GeV, where
H1 points are lower by about two standard deviations.

% Non-BibTeX users please use


\begin{thebibliography}{}
%
%%%%%%%%
% DVCS %
%%%%%%%%
\bibitem{h1-dvcs}
%\bibitem{Adloff:2001cn}
C.~Adloff {\it et al.}  [H1 Collaboration],
%``Measurement of deeply virtual Compton scattering at HERA,''
 \Journal{\PLB}{517}{2001}{47},
[hep-ex/0107005].
%%CITATION = HEP-EX 0107005;%%

\bibitem{zeus-dvcs}
%\bibitem{:2003ya}
 ZEUS Collaboration,
%``Measurement of deeply virtual Compton scattering at HERA,''
 \Journal{\PLB}{573}{2003}{46-62},
%DESY-03-059, 
[hep-ex/0305028].

\bibitem{h1eps03}
H1 Collaboration, contributed paper 115 to EPS03, Aachen.
 
% QCD Predictions
%%%%%%%%%%%%%%%%%%

\bibitem{Freund:2001hm}
 A.~Freund and M.~F.~McDermott,
 %``A next-to-leading order analysis of Deeply Virtual Compton
 %Scattering,''
 \Journal{\PRD}{65}{2002}{091901},
 [hep-ph/0106124].
 
\bibitem{Freund:2001hd}
 A.~Freund and M.~McDermott,
 % A detailed next-to-leading order QCD analysis of deeply virtual
 % Compton
 % scattering observables
 \Journal{\EPJ}{23}{2002}{651-674},
 [hep-ph/0111472].

\bibitem{Freund:2002qf}
A.~Freund, M.~McDermott and M.~Strikman,
%``Modelling generalized parton distributions to describe deeply virtual
%Compton scattering data,''
 \Journal{\PRD}{67}{2003}{036001},
[hep-ph/0208160].
%%CITATION = HEP-PH 0208160;%%

% Dipole Models
%%%%%%%%%%%%%%%%

\bibitem{Favart:2003cu}
 L.~Favart and M.~V.~Machado,
 %``Deeply virtual Compton scattering and saturation approach,''
 Eur.\ Phys.\ J.\ B{\bf C29}, 365-371 (2003),
 [hep-ph/0302079].

\bibitem{Donnachie:2000px}
 A.~Donnachie and H.~G.~Dosch,
 %``Diffractive exclusive photon production in deep inelastic
 %scattering,''
  Phys.\ Lett.\ B{\bf 502} (2001) 74-78,
 [hep-ph/0010227].
 %%CITATION = HEP-PH 0010227;%%
 
%\cite{Golec-Biernat:1999qd}
 \bibitem{Golec-Biernat:1999qd}
 K.~Golec-Biernat and M.~Wusthoff,
 %``Saturation in diffractive deep inelastic scattering,''
 Phys.\ Rev.\ D {\bf 60} (1999) 114023,
 [hep-ph/9903358].
 %%CITATION = HEP-PH 9903358;%%

%\cite{Favart:2004uv}
\bibitem{Favart:2004uv}
 L.~Favart and M.~V.~T.~Machado,
 %``QCD evolution and skewedness effects in color dipole description of
 %DVCS,''
 Eur.\ Phys.\ J.\ C {\bf 34} (2004) 429
 [hep-ph/0402018].
 %%CITATION = HEP-PH 0402018;%%

%\cite{Bartels:2002es}
 \bibitem{Bartels:2002es}
 J.~Bartels, K.~Golec-Biernat and H.~Kowalski,
 %``DGLAP evolution in the saturation model,''
 Acta Phys.\ Polon.\ B {\bf 33} (2002) 2853,
 [hep-ph/0207031].
 %%CITATION = HEP-PH 0207031;%%
\end{thebibliography}
\end{document}